\documentclass[conference]{IEEEtran}
\IEEEoverridecommandlockouts

\usepackage{cite}
\usepackage{amsmath,amssymb,amsfonts}
\usepackage{graphicx}
\usepackage{textcomp}
\usepackage{xcolor}
\usepackage{url}
\usepackage[hidelinks]{hyperref}
\hypersetup{
  pdftitle={From Capability to Assurance in Autonomous Penetration-Testing Harnesses: A Framework and Reference Implementation},
  pdfauthor={Joas Antonio dos Santos},
  pdfsubject={Autonomous penetration testing; assurance properties for LLM-agent harnesses},
  pdfkeywords={autonomous penetration testing, LLM agents, assurance, evidence grounding, capability-based authorization, tamper-evident logging, CVSS, offensive security}
}
\usepackage{booktabs}
\usepackage{multirow}
\usepackage{listings}
\usepackage{tikz}
\usetikzlibrary{shapes.geometric, arrows.meta, positioning, fit, backgrounds, calc}
\usepackage{float}

\newcommand{\cmark}{\checkmark}
\newcommand{\pmark}{$\sim$}
\newcommand{\xmark}{---}

\begin{document}

\title{From Capability to Assurance in Autonomous\\ Penetration-Testing Harnesses:\\ A Framework and Reference Implementation}

\author{\IEEEauthorblockN{Joas Antonio dos Santos}
\IEEEauthorblockA{\textit{Independent Researcher --- AI and Offensive Security} \\
S\~ao Paulo, SP, Brazil \\
joas.santos@redteamleaders.com}
}

\maketitle

\begin{abstract}
Research on large-language-model (LLM) agents for penetration testing is evaluated almost entirely by capability: whether the agent captures a flag or reproduces a proof-of-concept. This metric is appropriate for a benchmark but silent on the properties that decide whether an autonomous agent can be used in an authorized engagement---whether a reported finding is \emph{true}, whether the agent remained inside its \emph{authorized} scope, and whether an operator can \emph{audit} what it did. We call these \emph{assurance} properties and argue that they belong to the \emph{harness}---the runtime wrapping the model---and can be enforced in code. This paper makes three contributions. First, we define a framework of five assurance properties (evidence grounding, non-destructive claim reduction, computed severity, enforced authorization, and tamper-evident accountability), each with a formal model and an explicit acceptance test, and we connect them to prior work in capability-based security, tamper-evident logging, and software provenance. Second, we position representative systems (PentestGPT, the Cochise reference harness, MAPTA, and the trajectory judge PentestJudge) within the framework using published, coded criteria, and identify a consistent assurance gap. Third, we study one open-source implementation, NeuroSploit (pinned to an exact commit), reporting its architecture, its complexity cost measured by documented commands, and a version-pinned, content-addressed artifact bundle from a run against a public deliberately-vulnerable target. We are precise about what is and is not evidenced: we execute the deterministic P4 and P5 acceptance tests directly and, in doing so, find and report a real enforcement gap, which we reflect by scoring both P4 and P5 as partial. We therefore claim an \emph{initial existence argument} that the properties are realizable together---not a comparative-performance result---and we specify the multi-target, ablation, and adversarial evaluation protocol required to turn the framework's obligations into measurements. The contribution is a vocabulary, a set of realizable design obligations, and a measurable evaluation plan for autonomous offensive agents whose output can withstand scrutiny.
\end{abstract}

\begin{IEEEkeywords}
autonomous penetration testing, large language models, assurance, evidence grounding, capability-based authorization, tamper-evident logging, CVSS, offensive security
\end{IEEEkeywords}

\section{Introduction}\label{sec:intro}

Autonomous penetration testing has become a measurable engineering discipline. Since Fang et al.\ showed that an LLM agent could exploit real one-day web vulnerabilities from a CVE description \cite{fang2024oneday}, systems such as PentestGPT \cite{deng2023pentestgpt}, HPTSA \cite{zhu2024hptsa}, MAPTA \cite{david2025mapta}, CHECKMATE \cite{wang2025checkmate}, Pentest-R1 \cite{kong2025pentestr1}, and the Cochise reference harness \cite{happe2026cochise} have advanced the state of the art, and benchmarks such as CyberGym \cite{wang2025cybergym} and ExploitGym \cite{exploitgym2026} now measure offensive capability across hundreds to thousands of real targets.

Almost all of this progress is measured with one family of metrics: task or subtask success, operationalized as capturing a flag or reproducing a proof-of-concept (PoC). That instrument is right for a benchmark. It is silent, however, on three properties that determine whether an autonomous agent can be used in a paid, authorized engagement:

\begin{itemize}
\item \textbf{Truth.} A flag capture is self-verifying; a client report is not. An agent that emits a fluent but unproven vulnerability narrative---the ``hallucinated compliance'' pattern of Nguyen and Husain \cite{nguyen2025agenticpentest}---imposes remediation cost and destroys credibility.
\item \textbf{Authorization.} The engagement boundary belongs to the client. If scope exists only as prompt text, an agent that follows an off-target redirect or acts on a discovered subdomain has crossed a legal line, discovered only by reading the report.
\item \textbf{Accountability.} Afterwards the operator must state what was done, to what, under whose authority, and with what result. Prose logs are lossy and reorderable.
\end{itemize}

These are \emph{assurance} questions, and our central claim is that they are properties of the \emph{harness}---the runtime that surrounds the model with a shell, tool wrappers, a token budget, a memory policy, and an output parser---not of the model's weights. The harness is where an assurance obligation can be enforced in code rather than requested in a prompt, and the same base model in two harnesses can differ by tens of percentage points on the same benchmark \cite{cybench2024,wang2025checkmate}, which already establishes the harness as the decisive engineering surface. The assurance properties themselves are not new inventions: they instantiate, for offensive agents, long-standing ideas from capability-based security \cite{dennis1966,millercapability,birgisson2014macaroons}, enforceable security policies \cite{schneider2000}, tamper-evident logging \cite{schneier1999,crosby2009,merkle1987}, and provenance \cite{torres2019intoto,provdm}. Our framework's role is to name which of them a pentest harness owes, and to state each as a testable obligation.

\textbf{Research questions.}
\begin{itemize}
\item[\textbf{RQ1}] What assurance properties should an autonomous penetration-testing harness provide, and can each be stated as an implementation-independent obligation with an explicit acceptance test?
\item[\textbf{RQ2}] Where do existing systems sit with respect to these properties under objective, coded criteria?
\item[\textbf{RQ3}] Are the properties realizable in a single system, at what complexity cost, and what evaluation would measure how completely each is satisfied?
\end{itemize}

\textbf{Contributions.}
\begin{enumerate}
\item A \emph{framework} of five assurance properties (Section~\ref{sec:framework}), each with a formal model and an acceptance test, related to prior security literature; this answers RQ1.
\item A \emph{positioning} of representative systems using published coding criteria (Section~\ref{sec:related} and Appendix~\ref{app:table}), answering RQ2.
\item A \emph{reference implementation study} of NeuroSploit \cite{neurosploit2026}, pinned to an exact commit, with complexity measured by documented commands and a content-addressed artifact bundle (Sections~\ref{sec:instance}--\ref{sec:eval}); and a \emph{measurable evaluation protocol}---multi-target precision/recall, ablation, and adversarial P4/P5 batteries (Appendix~\ref{app:protocol})---that specifies what remains to be measured. Together these answer RQ3.
\end{enumerate}

We state scope precisely. We make \emph{no} comparative-performance claim: no shared benchmark for assurance properties exists, and constructing one is the central open problem (Section~\ref{sec:threats}). The reference study is an \emph{initial existence argument}, not a proof that every property holds on every action; Section~\ref{sec:eval} is explicit about what the archived artifacts do and do not evidence.

\section{Background and Positioning}\label{sec:related}

\subsection{Autonomous offensive agents and their harnesses}
Early autonomous LLM pentesters were single-agent ReAct loops \cite{fang2024websites,fang2024oneday}. Multi-agent designs followed: HPTSA showed a planner dispatching task-specific subagents outperforms a generalist on zero-day-class web targets \cite{zhu2024hptsa}, and the Planner--Executor--Perceptor pattern was formalized in CHECKMATE, which, holding the base model fixed, obtains a $>20\%$ success improvement and $>50\%$ cost reduction from the harness alone \cite{wang2025checkmate}. MAPTA added tool-grounded execution with end-to-end exploit validation \cite{david2025mapta}. Cochise \cite{happe2026cochise} is the closest work in intent: a deliberately small (630 LOC) Python reference harness for reproducible \emph{traces}. The properties we study concern assurance of \emph{output}.

\subsection{Evaluation and judging}
Cybench evaluates agents on 40 professional CTF tasks and is reported in most frontier-model system cards \cite{cybench2024}. PentestEval shows end-to-end pipeline success collapsing to ${\le}31\%$ even when per-stage scores are high \cite{yang2025pentesteval}. CyberGym \cite{wang2025cybergym} and ExploitGym \cite{exploitgym2026} scale evaluation to real code. PentestJudge \cite{caldwell2025pentestjudge} argues a flag-only reward is gameable and grades agent \emph{behavior}; that is an assurance argument at the trajectory level, which our framework complements at the level of findings, authorization, and per-action accountability.

\subsection{Security foundations}
Each property below has a lineage. P4 is a capability \cite{dennis1966,millercapability}; the concrete token design follows Macaroons \cite{birgisson2014macaroons} and JWT \cite{rfc7519} with HMAC \cite{rfc2104}. P5 is tamper-evident logging \cite{schneier1999,crosby2009} over a hash chain \cite{merkle1987}, whose full-rewrite weakness is addressed by external anchoring as in Certificate Transparency \cite{rfc6962} and supply-chain attestation \cite{torres2019intoto}. P1 relates to grounding and attribution for LLM output \cite{gao2023rarr,manakul2023selfcheckgpt} and to prompt-injection defenses for tool-using agents \cite{greshake2023injection,yang2025mcpsecbench}. P3 builds on CVSS v3.1 \cite{first2019cvss}, and the enforcement stance follows Schneider's enforceable security policies \cite{schneider2000} and W3C provenance \cite{provdm}.

\subsection{The assurance gap}
Appendix~\ref{app:table} states objective coding criteria and applies them, per analyzed version, to representative systems; Table~\ref{tab:related} summarizes. The pattern is consistent: capability is addressed and, at best, trajectory-level judging; finding-level truth, enforced scope, and per-action accountability are left to the surrounding process. This gap---not any single missing feature---is what the framework names.

\begin{table}[t]
\centering
\caption{Assurance properties (P1--P5) across representative systems, coded per Appendix~\ref{app:table}. \cmark{}=addressed in the analyzed artifact; \pmark{}=partial/process-level; \xmark{}=not addressed in the analyzed artifact (absence of a claim, not proof of absence in code). NeuroSploit P4/P5 are \pmark{} because the primary-target check (P4) and external anchoring (P5) are not yet in the tested build; see Sections~\ref{sec:accept} and~\ref{sec:threats}.}
\label{tab:related}
\resizebox{\columnwidth}{!}{%
\begin{tabular}{@{}lccccc@{}}
\toprule
 & \textbf{PentestGPT} & \textbf{Cochise} & \textbf{MAPTA} & \textbf{PentestJudge} & \textbf{NeuroSploit} \\
 & \cite{deng2023pentestgpt} & \cite{happe2026cochise} & \cite{david2025mapta} & \cite{caldwell2025pentestjudge} & \cite{neurosploit2026} \\
\midrule
P1 Evidence grounding & \xmark & \xmark & \cmark & \pmark & \cmark \\
P2 Non-destr.\ reduction & \xmark & \xmark & \pmark & \pmark & \cmark \\
P3 Computed severity & \xmark & \xmark & \xmark & \xmark & \pmark \\
P4 Enforced authorization & \xmark & \xmark & \xmark & \xmark & \pmark \\
P5 Tamper-evident audit & \xmark & \xmark & \xmark & \xmark & \pmark \\
\midrule
Trajectory judging & \xmark & \pmark & \xmark & \cmark & \pmark \\
Released implementation & \cmark & \cmark & \cmark & \cmark & \cmark \\
\bottomrule
\end{tabular}%
}
\end{table}

\section{Threat Model}\label{sec:threat}

\textbf{Setting.} An operator runs a harness against systems for which a client has granted written authorization. Operator and authorizing party may differ. The target is partially observable and may be hostile to the agent.

\textbf{Assets.} (A1) client systems and data, including assets adjacent to but outside scope; (A2) report integrity---no false findings, no silently dropped real ones; (A3) the operator's ability to prove, afterwards, that the engagement stayed within authorization.

\textbf{Adversaries and failure modes.} (T1) the \emph{model} as a source of confident hallucination; (T2) a \emph{hostile target} performing prompt injection through tool output \cite{greshake2023injection,yang2025mcpsecbench}; (T3) \emph{scope drift} onto discovered-but-unauthorized assets, including redirects, DNS rebinding, alternate IP encodings, and mid-run DNS changes; (T4) an \emph{after-the-fact dispute}, including an adversary who can edit, remove, reorder, or wholesale-rewrite the log. The framework addresses T1, T3, and T4 structurally; T2 is reduced but not eliminated (Section~\ref{sec:threats}). For T4 we distinguish a \emph{local} adversary (can edit files but not forge external signatures) from a \emph{full-rewrite} adversary (can recompute an unanchored chain); P5's acceptance test targets both only when external anchoring is present (Appendix~\ref{app:audit}).

\section{A Framework of Assurance Properties}\label{sec:framework}

We state five properties as obligations on the harness, each with a formal model and an \emph{acceptance test}: a concrete check that decides whether an implementation satisfies the property. A \emph{finding} $f$ is a claim about a target; the \emph{evidence ledger} $E=\{e_1,\dots,e_m\}$ is the set of receipts collected during an engagement (raw tool output: an HTTP response, an out-of-band callback, an error oracle, a shell transcript, or, white-box, a source citation). Each receipt $e$ records a target, an action reference, a timestamp, and a content hash.

\subsection{P1 --- Evidence grounding}
\emph{No claim enters the reported world model without a receipt that a decision procedure outside the model accepts.}
Let $\mathrm{cite}(c)\subseteq E$ be the receipts a claim $c$ names. Define the external decision procedure $\mathrm{supports}$ (formalized in Appendix~\ref{app:supports}) returning a verdict in $\{\textsf{Confirmed},\textsf{Rejected},\textsf{NeedsReview}\}$. P1 requires
\begin{equation}
\forall c\in f:\ \mathrm{cite}(c)\neq\emptyset \ \wedge\ \mathrm{supports}(c,\mathrm{cite}(c))\neq\textsf{Rejected},
\end{equation}
with any \textsf{NeedsReview} claim labelled and withheld from asserted status. The obligation is that $\mathrm{supports}$ is \emph{not} ``an LLM agrees''; model agreement is not evidence.
\emph{Acceptance test (T-P1):} inject a syntactically plausible but fabricated receipt (a transcript with no matching action hash); the finding must not reach asserted status.

\subsection{P2 --- Non-destructive claim reduction}
\emph{An overstated finding is reduced to what its evidence supports, never silently deleted or inflated, and a claim with no security-relevant mechanic is not force-retained.}
Decompose $f$ into separable claims and partition them into supported mechanic $M(f)$ and unsupported impact $I(f)$. The reduction returns one of four typed outcomes (Appendix~\ref{app:prosecutor}): \textsf{supported\_finding} ($M(f)$ rewritten), \textsf{informational\_observation} (a benign but true fact), \textsf{insufficient\_evidence}, or \textsf{invalid\_claim}. Formally,
\begin{equation}
f'=
\begin{cases}
\mathrm{rewrite}(M(f)) & M(f)\ \text{security-relevant},\\
\mathrm{demote}(M(f)) & M(f)\ \text{true but benign},\\
\emptyset & \text{otherwise.}
\end{cases}
\end{equation}
\emph{Acceptance test (T-P2):} a finding whose impact is unproven but whose mechanic is proven (e.g., a measured missing rate limit) must be retained and downgraded, not rejected; a finding with neither must not be force-retained.

\subsection{P3 --- Computed, evidence-graded, impact-separated severity}
\emph{Severity is computed from evidence by a deterministic function and capped by demonstrated impact; potential impact is recorded separately so that safely-unexploited worst cases are not understated.}
Let $g$ be the CVSS v3.1 base function \cite{first2019cvss} over a full metric vector whose values each cite a receipt; let $\ell_d(f)$ be the \emph{demonstrated}-impact rung and $\ell_p(f)$ the \emph{potential}-impact rung on
\begin{equation}
\text{reached} < \text{read} < \text{wrote} < \text{RCE} < \text{crossed}.
\label{eq:ladder}
\end{equation}
P3 requires a demonstrated score and a separately-recorded potential score,
\begin{equation}
\mathrm{sev}_d(f)=\min\!\big(g(f),\mathrm{cap}(\ell_d(f))\big),\quad
\mathrm{sev}_p(f)=g(f),
\end{equation}
each annotated with an evidence-confidence level. The separation prevents the failure the reviewer of an authorized test rightly fears: correctly declining to demonstrate destructive impact must not silently lower the recorded risk. Appendix~\ref{app:cvss} gives the cap table and a worked example.
\emph{Acceptance test (T-P3):} two findings of the same CWE with different demonstrated impact must receive different $\mathrm{sev}_d$; class-only scoring fails.

\subsection{P4 --- Enforced, attributable authorization}
\emph{Scope is a signed grant, bound to a client, an operator, an engagement, and an asset set, that in-session behavior can narrow but never widen, with replay and revocation controls.}
Let $A$ be an authorization token carrying issuer (client key id), audience (harness instance), engagement id, operator id, scope (hosts, CIDRs, URL prefixes), action ceiling, environment, validity window $[\mathit{nbf},\mathit{exp}]$, and nonce; let $\sigma$ be its signature. P4 requires that an action $a$ on target $t$ be admitted only if
\begin{equation}
\begin{aligned}
&\mathrm{verify}(A,\sigma,K_{\mathrm{iss}})\ \wedge\ \mathit{nbf}\le \mathit{now}\le \mathit{exp}\ \wedge\ \mathrm{fresh}(\mathrm{nonce})\\
&\wedge\ \mathrm{aud}(A)=\mathrm{self}\ \wedge\ (t,a)\sqsubseteq A\ \wedge\ (t,a)\sqsubseteq\mathrm{cfg},
\end{aligned}
\end{equation}
where $\mathrm{cfg}$ is local configuration and $\sqsubseteq$ is ``within.'' $A$ is a ceiling: no in-session reasoning can yield $(t,a)\not\sqsubseteq A$. Appendix~\ref{app:token} specifies the token, key custody, revocation, and rotation, and notes that HMAC \cite{rfc2104} proves only holder-of-key, so separating client from operator argues for an asymmetric signature (client-held private key, harness-held public key), which we present as the recommended evolution of the current HMAC form.
\emph{Acceptance test (T-P4):} the adversarial scope battery of Appendix~\ref{app:protocol} (off-scope redirect, DNS rebinding, alternate IP encodings, userinfo URLs, over-broad wildcard, expired/tampered token, in-session widening) must each be denied.

\subsection{P5 --- Externally-anchored tamper-evident accountability}
\emph{Every action is recorded in an append-only log whose integrity is verifiable, and whose root is anchored outside the writing process so that a full rewrite is detectable.}
Actions produce records $r_1,\dots,r_n$ chained by $h_i=H(r_i\,\|\,h_{i-1})$, $h_0=H(\text{genesis})$. A hash chain alone detects only local edits; a full-rewrite adversary can recompute it. P5 therefore additionally requires periodic anchoring of $h_i$---an external signature, a trusted timestamp, a write-once store, or publication to a transparency log \cite{rfc6962,crosby2009}---so that no post-hoc chain can match the anchored roots.
\emph{Acceptance test (T-P5):} the tamper battery of Appendix~\ref{app:protocol} (record removed, altered, reordered; evidence substituted; token swapped; log truncated; whole-chain rebuilt) must each be detected; the last only when anchoring is enabled.

\medskip
\noindent\textbf{Relationships and completeness.} P1 is prerequisite to P2 and P3. P4 and P5 are independent of P1--P3 and of the model. The five are proposed as necessary for an engagement that must withstand scrutiny; they are not claimed exhaustive---trajectory-level judging \cite{caldwell2025pentestjudge} is a candidate sixth, kept complementary because it ranges over runs rather than findings or actions.

\section{Reference Implementation}\label{sec:instance}

To address RQ3 we study \textbf{NeuroSploit} \cite{neurosploit2026}, an open-source Rust harness developed by the author (disclosed as a competing interest, Section~\ref{sec:ethics}). It is used as an existence argument and a source of complexity measurements, not a product claim; any harness satisfying P1--P5 would serve. All measurements in this paper are pinned to one commit (Table~\ref{tab:pin}) and produced by the commands in Appendix~\ref{app:repro}.

\begin{table}[h]
\centering
\caption{Evaluated version (all Section~\ref{sec:impl}--\ref{sec:eval} numbers refer to this state).}
\label{tab:pin}
\resizebox{\columnwidth}{!}{%
\begin{tabular}{@{}ll@{}}
\toprule
\textbf{Item} & \textbf{Value} \\
\midrule
Commit SHA & \texttt{8894649ccb42a04c4763085b3b663e49fd98ddf5} \\
Describe/tag & \texttt{v4.0.0-32-g8894649} \\
Commit date & 2026-09-18 \\
Measurement date & 2026-09-18 \\
OS & macOS 15.1 (Darwin 24.1.0, arm64) \\
Toolchain & \texttt{rustc}/\texttt{cargo} 1.96.0 \\
\bottomrule
\end{tabular}}
\end{table}

\begin{figure}[t]
\centering
\resizebox{\columnwidth}{!}{%
\begin{tikzpicture}[
    node distance=0.55cm and 0.5cm,
    stage/.style={rectangle, draw, rounded corners=2pt, fill=blue!10, text width=2.5cm, minimum height=0.8cm, align=center, font=\scriptsize},
    wrap/.style={rectangle, draw, dashed, rounded corners=3pt, fill=red!5, font=\scriptsize\itshape, inner sep=4pt},
    ar/.style={-{Stealth[length=2mm]}, thick},
]
\node[stage] (recon) {Recon \& belief state};
\node[stage, right=of recon] (sel) {Agent selection (surface-matched)};
\node[stage, right=of sel] (run) {Parallel run over model pool};
\node[stage, below=of run] (ver) {Verification (P1--P3)};
\node[stage, below=of sel] (chain) {Chaining \& attack graph};
\node[stage, below=of recon] (rep) {Evidence-graded report};
\draw[ar] (recon)--(sel);
\draw[ar] (sel)--(run);
\draw[ar] (run)--(ver);
\draw[ar] (ver)--(chain);
\draw[ar] (chain)--(rep);
\begin{scope}[on background layer]
\node[wrap, fit=(recon)(sel)(run)(ver)(chain)(rep), label={[font=\scriptsize\itshape]above:P4 authorization ceiling $\cdot$ P5 hash-chained audit (every action)}] {};
\end{scope}
\end{tikzpicture}%
}
\caption{NeuroSploit pipeline. Recon/select/run/verify/chain/report are wrapped by the P4 ceiling and the P5 chain.}
\label{fig:arch}
\end{figure}
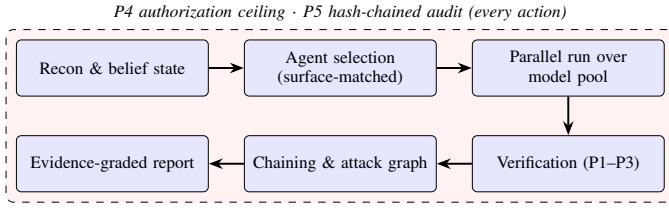

Figure~\ref{fig:arch} shows the pipeline: recon builds a belief state; a selector chooses surface-matched agents from a markdown knowledge base; agents run in parallel over a model pool; outputs pass the verification pipeline; survivors are chained and reported. The P4 authorization ceiling and the P5 audit chain wrap every action.

\subsection{Planning as an anti-hallucination gate}
The target is partially observable, so the harness tracks a \emph{belief}: a property graph whose nodes carry a probability and an evidence count, updated from observations by a Bayesian step; per-node Shannon entropy $H(b)$ measures diffuseness. The ``recon vs.\ exploit'' choice is a value-of-information decision: with $V_{\text{obs}}(b)$ and $V_{\text{exploit}}(b)$ the expected payoffs, the planner observes while
\begin{equation}
V_{\text{obs}}(b) > V_{\text{exploit}}(b),
\label{eq:voi}
\end{equation}
which holds when the belief is diffuse; the same inequality bars asserting exploitability while $H(b)$ is high, reinforcing P1 at the planning layer.

\subsection{Realizing P1--P3}
\begin{figure}[t]
\centering
\resizebox{\columnwidth}{!}{%
\begin{tikzpicture}[
    node distance=0.4cm,
    vrow/.style={rectangle, draw, rounded corners=2pt, fill=green!10, text width=7.6cm, minimum height=0.6cm, align=left, font=\scriptsize},
]
\node[vrow] (g) {\textbf{1. Grounding (P1).} Receipt required; ungrounded $\Rightarrow$ withheld.};
\node[vrow, below=of g] (d) {\textbf{2. Deterministic validator (P1).} Per-CWE, outside the model.};
\node[vrow, below=of d] (v) {\textbf{3. Adversarial vote.} $N$-model agreement + refute, over survivors.};
\node[vrow, below=of v] (p) {\textbf{4. Prosecutor (P2).} Four typed outcomes; rewrites, no silent delete.};
\draw[-{Stealth[length=2mm]}, thick] (g)--(d);
\draw[-{Stealth[length=2mm]}, thick] (d)--(v);
\draw[-{Stealth[length=2mm]}, thick] (v)--(p);
\end{tikzpicture}%
}
\caption{Verification pipeline realizing P1--P2; severity (P3) follows.}
\label{fig:verify}
\end{figure}
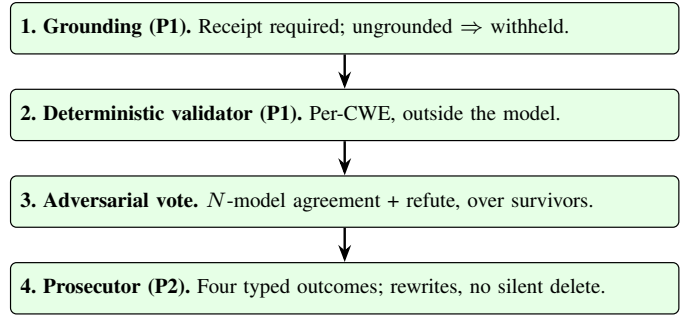

Grounding (P1) enforces a receipt in the engagement-appropriate mode (empirical / symbolic / either). Deterministic per-CWE validators (Appendix~\ref{app:validators}) implement $\mathrm{supports}$ outside the model: 27 validators, keyed to CWE classes and referencing 31 CWE identifiers, each returning \textsf{Confirmed}/\textsf{Rejected}/\textsf{NeedsReview}. Adversarial $N$-model voting runs only over survivors. The evidence prosecutor realizes P2 with the four typed outcomes above. Severity (P3) is then computed and impact-separated (Appendix~\ref{app:cvss}). Figure~\ref{fig:verify} shows the order; deterministic checks precede model-based ones.

\subsection{Realizing P4--P5}
Authorization is a signed capability token (Appendix~\ref{app:token}) verified as the P4 ceiling, with a two-layer scope guard: hard scope (allowlist defaulting to the engagement target, so discovery cannot widen scope) and soft scope (read-only zones, forbidden methods, rate ceilings). Every action emits a hash-chained record (Appendix~\ref{app:audit}) with a \texttt{verify()} routine; external anchoring is discussed there as the defense against full rewrite. An additive OT risk model $r_{\text{eff}}=r_{\text{action}}+c_{\text{asset}}+r_{\text{proto}}+p_{\text{priv}}+b_{\text{radius}}$ forbids disruptive industrial function codes while permitting the reads OT findings come from. These mechanisms are exercised both by the unit-test suite (Section~\ref{sec:impl}) and by a command-line acceptance battery we ran directly (Section~\ref{sec:accept}); the latter also surfaced a real enforcement gap, reported below.

\subsection{Executed acceptance tests for P4 and P5}\label{sec:accept}
Because the P4 and P5 acceptance tests are deterministic and need no model, we executed them on the pinned build (Table~\ref{tab:pin}). Table~\ref{tab:accept} reports the P4 capability-token battery. Token issuance, verification, tamper rejection, wrong-key rejection, and expiry detection all behave as specified, and \texttt{constrain()} correctly drops an out-of-grant host from local scope (the audit record for an off-grant run shows the hard scope reduced to the granted host). \emph{However}, when the run's \emph{primary target} lies outside the grant, the offline engagement still reconned it and logged \texttt{engagement-start} as \texttt{allow}: the token constrains the agent scope but the tested build does not refuse an out-of-grant primary target at engagement start. We therefore score P4 as \emph{partial} (Table~\ref{tab:related}) and list this as a limitation (Section~\ref{sec:threats}). For P5, an offline run produced a hash-chained \texttt{audit.jsonl} (fields \texttt{seq}, \texttt{prev\_hash}, \texttt{hash}$=\mathrm{sha256}(\text{digest})$), and the audit module's tests---\texttt{the\_chain\_verifies\_and\_notices\_tampering}, \texttt{removing\_a\_record\_breaks\_the\_chain\_too}, \texttt{reopening\_continues\_the\_same\_chain}, \texttt{refusals\_are\_recorded\_not\_just\_actions}---pass, covering the alter/remove/reorder cases of T-P5; external anchoring against a full-rewrite adversary remains design-only. Of the harness's 312 passing unit tests, 109 exercise the P1--P5 modules directly---P1 (grounding, validation) 52, P2 (claims, prosecutor) 19, P4 (capability, scope) 27, P5 (audit) 11---while the remainder cover recon, planning, chaining, and reporting; P3's severity-cap logic is covered within the P1 validation and P2 claim tests rather than a dedicated module. The full suite reports \texttt{312 passed} at the pinned commit.

\begin{table}[h]
\centering
\caption{Executed P4 capability-token battery (CLI, deterministic).}
\label{tab:accept}
\resizebox{\columnwidth}{!}{%
\begin{tabular}{@{}lll@{}}
\toprule
\textbf{Case} & \textbf{Expected} & \textbf{Observed} \\
\midrule
Valid token & verify & verified, grant printed \\
Tampered token & refuse & refused (signature) \\
Wrong signing key & refuse & refused \\
Expired ($\mathit{exp}$ past) & refuse & flagged expired \\
Off-grant host in local scope & drop & dropped from hard scope \\
Off-grant \emph{primary target} & refuse & \textbf{reconned (gap)} \\
\bottomrule
\end{tabular}%
}
\end{table}

\section{Implementation Cost}\label{sec:impl}

Table~\ref{tab:impl} reports the pinned instance, measured by Appendix~\ref{app:repro}. Enforcing P1--P5 is not free: the validators, token and audit machinery, and multi-model verification account for a substantial fraction of the code, and the multi-model vote adds per-finding token and latency cost over a single-model pipeline (quantified by the protocol of Appendix~\ref{app:protocol}). The scale itself is a data point for RQ3: assurance is realizable at tens of modules and tens of thousands of lines, an order of magnitude beyond a trace-reproduction harness such as Cochise (630 LOC \cite{happe2026cochise}), which is consistent with the gap being real engineering rather than a prompt.

\begin{table}[h]
\centering
\caption{Complexity of the pinned instance (commands in Appendix~\ref{app:repro}).}
\label{tab:impl}
\small
\begin{tabular}{@{}lr@{}}
\toprule
\textbf{Component} & \textbf{Metric} \\
\midrule
Rust modules (harness crate) & 40 \\
Rust LOC (harness crate) & 23{,}370 \\
Rust LOC (workspace, excl.\ \texttt{target}) & 28{,}802 \\
Automated tests (\texttt{\#[test]}/\texttt{\#[tokio::test]}) & 312 \\
Deterministic validators (impl \texttt{CweValidator}) & 27 \\
CWE identifiers referenced by validators & 31 \\
Agent definitions (markdown) & 446 \\
Supported model providers & 18 \\
\bottomrule
\end{tabular}
\end{table}

\section{Reference Run and Artifact Bundle}\label{sec:eval}

To characterize behavior---not to claim a comparative result---we examine an archived black-box run against \texttt{http://testaspnet.vulnweb.com/}, a public deliberately-vulnerable application published by Acunetix for testing (an ASP.NET/IIS~8.5 site; recon recorded zero security headers and a non-\texttt{Secure} \texttt{ASP.NET\_SessionId} cookie). The run scheduled 245 surface-matched agents and produced 27 findings after verification, from 25 distinct agents, spanning 13 CWE classes, in 3.55 hours, with 32 PoC scripts and 27 evidence screenshots (Table~\ref{tab:casestudy}).

\begin{table}[h]
\centering
\caption{Archived reference run (\texttt{testaspnet.vulnweb.com}, black-box).}
\label{tab:casestudy}
\small
\begin{tabular}{@{}lr@{}}
\toprule
\textbf{Quantity} & \textbf{Value} \\
\midrule
Agents scheduled & 245 \\
Agents producing a verified finding & 25 \\
Findings after verification & 27 \\
\quad Critical / High / Medium / Low / Info & 1 / 6 / 14 / 4 / 2 \\
Distinct CWE classes & 13 \\
PoC scripts / evidence screenshots & 32 / 27 \\
Wall-clock time & 3.55\,h \\
Artifact bundle root (SHA-256) & \texttt{5e018a1e\ldots55bce3} \\
\bottomrule
\end{tabular}
\end{table}

\textbf{What the archived run evidences, precisely.} The artifacts record, per finding, the receipt text, PoC reference, adversarial \texttt{votes}, a \texttt{validated} flag, a \texttt{review\_status}, and a severity \emph{label} (P1, P2, and the labelling half of P3). They do \emph{not} contain populated CVSS \emph{vectors} (the \texttt{cvss} field is empty for all 27), nor a capability token, nor an audit-log file: this run predates full instrumentation of those outputs. CVSS-vector computation (P3) is evidenced by the code and its 312 unit tests, not by this archived run. Token verification (P4) and the hash-chained audit log (P5) are evidenced separately by the executed acceptance battery of Section~\ref{sec:accept}, which produced a real capability token and a hash-chained \texttt{audit.jsonl} and also revealed the P4 primary-target gap. The archived target run itself contains no audit log; the bundle additionally contains a separate \texttt{audit.jsonl} produced by the deterministic P5 acceptance run (Section~\ref{sec:accept}) and must not be read as the audit trail of the archived engagement. The archived run's model version, temperature, vote count, token budget, concurrency, and seed were not persisted and are reported as \emph{unknown}; the harness default at that time was \texttt{anthropic:claude-opus-4-8}. Producing a single fully-instrumented \emph{engagement} run that carries all five properties' artifacts end-to-end---with those parameters recorded---is part of the protocol in Appendix~\ref{app:protocol}.

\textbf{Reproducibility.} The non-sensitive artifacts (recon, findings, PoCs, exploitation notes, the executed acceptance results, and the produced \texttt{audit.jsonl}; the public target contains no third-party data) form a \emph{content-addressed} bundle \texttt{assurance-artifacts-v1.tar.gz} (40 files, 84~KB, MIT-licensed), available from the project repository \cite{neurosploit2026} at the pinned commit (Table~\ref{tab:pin}), with per-file SHA-256 manifest root
\begin{center}\ttfamily\scriptsize 5e018a1e1a9dc2d19dd82eb0fb3e8af2\\7351357f72932d7e7a7df5ae8655bce3\end{center}
The published hash is a content identifier for the bundle, \emph{not} an external anchor for the audit chain (Appendix~\ref{app:audit}). Appendix~\ref{app:repro} lists the commands to recompute every number and the manifest root, and the pinned commit (Table~\ref{tab:pin}) fixes the code.

\textbf{An honest negative.} On separate runs against \texttt{testphp.vulnweb.com} (a PHP/MySQL sibling known to contain SQL injection), the instance scheduled 245 agents but reported \emph{zero} confirmed findings. We report this. The deterministic SQLi validator (CWE-89) requires a reproducible baseline-versus-attack differential; those runs terminated at the observation phase---consistent with Equation~\eqref{eq:voi}---before an agent produced one. The behavior is correct by P1 yet is a false \emph{negative} against a truly vulnerable target, illustrating that a strict evidence rule trades recall for precision. The protocol of Appendix~\ref{app:protocol} is designed to quantify exactly this trade-off.

\section{Discussion, Limitations, and Threats to Validity}\label{sec:threats}

\textbf{No assurance benchmark; single instrumented run (primary threats).} No shared, public benchmark measures truth-under-scrutiny, scope enforcement, or auditability, so every ranking claim is out of scope, and the empirical evidence here is one archived behavioral run plus a unit-tested codebase. The paper's claim is an \emph{initial existence argument}, not a demonstration that each property holds on every action. Appendix~\ref{app:protocol} specifies the missing measurements: at least three to five vulnerable applications with known ground truth (e.g., OWASP Juice Shop, WebGoat, DVWA at fixed versions, a deliberately-vulnerable API, one authorized real scope), $\ge 3$ runs per target, and reported true/false positives and negatives, precision, recall, cost, tokens, and time; an ablation removing each of P1's deterministic validators, the vote, the prosecutor, and the P3 cap, and comparing common logging against P5; adversarial P4 and P5 batteries; and baselines (NeuroSploit with P1--P5 disabled, a plain ReAct loop, a single-model configuration, and a conventional scanner on the same targets).

\textbf{Possible memorization.} The archived run targets a widely-used public application that may appear in training data, which could inflate recall; the honest negative partially counters an over-optimistic reading, but external validity is limited until diverse, partly-private targets are used.

\textbf{Precision--recall.} P1's outside-the-model rule maximizes precision and can suppress true positives (the \texttt{testphp} negative); whether the resulting report is more useful than a higher-recall one is the empirical question the protocol answers.

\textbf{Residual surface (T2).} No harness eliminates prompt injection through hostile tool output; adversarial content can influence an agent's next step before verification runs. P1 limits the \emph{consequences}---an injected instruction cannot manufacture a receipt that passes a per-CWE validator---but the planning layer remains a target, and confining every agent-authored command is only partial in the current implementation.

\textbf{P4 primary-target gap (found by executing T-P4).} The acceptance battery of Section~\ref{sec:accept} revealed that, while the token correctly constrains agent scope, an offline engagement whose \emph{primary target} lies outside the grant is reconned and logged \texttt{allow} rather than refused. This is a concrete enforcement gap between the specification of P4 and the tested build; it is the reason P4 is scored partial, and closing it (checking the primary target against the constrained scope at engagement start) is the first fix the reference implementation owes.

\textbf{Property strength.} As specified, P4's HMAC form proves holder-of-key, not client intent; the asymmetric evolution (Appendix~\ref{app:token}) is required to bind client and operator cryptographically. P5's guarantee against a full-rewrite adversary holds only with external anchoring enabled (Appendix~\ref{app:audit}); without it the mechanism detects local, partial tampering only. We state these as scoped guarantees, not absolutes.

\section{Ethics and Responsible Use}\label{sec:ethics}

All runs discussed were conducted against targets published by their owners for security testing (the Acunetix \texttt{vulnweb.com} family); no third-party data, credentials, or personal information are disclosed, and evidence contains only synthetic markers chosen by the harness. Autonomous offensive tooling is dual-use; the assurance properties defined here---enforced authorization (P4), tamper-evident accountability (P5), and the OT safety model---are precisely the mechanisms that constrain misuse and make an authorized engagement defensible. We describe an architecture and release no exploit payloads for unpatched third-party vulnerabilities.

\textbf{Competing interests and funding.} The author is the developer of NeuroSploit, the reference instance; this is the paper's sole competing interest and the reason the study is framed as an existence argument rather than a comparison. The work received no external funding.

\textbf{Availability.} The implementation is open source \cite{neurosploit2026}, pinned by Table~\ref{tab:pin}; the run artifact bundle is anchored by the manifest root in Section~\ref{sec:eval}, enabling independent inspection.

\section{Conclusion}\label{sec:conclusion}

The useful frontier for autonomous penetration testing is shifting from capability to \emph{assurance}, and assurance is a property of the harness. We defined five implementation-independent properties, each with a formal model and an acceptance test, related them to capability-based security, tamper-evident logging, and provenance, positioned existing systems against them, and studied one open-source implementation as a version-pinned, content-addressed existence argument---while stating precisely what its archived run does and does not evidence and specifying the multi-target, ablation, and adversarial protocol that would turn the framework's obligations into measurements. Building that assurance benchmark is the work we hope this framework enables.

\section*{Acknowledgment}
The author thanks the maintainers of the public vulnerable-target and benchmark ecosystems. Generative language-model tooling assisted with drafting, reference verification, and copy-editing; all sources were checked by the author against their original records, and the author takes full responsibility for the content, including any errors.

\appendices

\section{Reproducibility Commands}\label{app:repro}
All numbers in Tables~\ref{tab:pin}--\ref{tab:casestudy} were produced at the pinned commit by the following, on the environment of Table~\ref{tab:pin}.
\begin{lstlisting}
# version
git rev-parse HEAD
git describe --tags --always
git log -1 --format=%ci
rustc --version

# complexity (Table III)
find agents_md -name '*.md' -type f | wc -l          # agents: 446
find crates/harness/src -name '*.rs' | wc -l         # modules: 40
find crates/harness/src -name '*.rs' -print0 \
  | xargs -0 cat | wc -l                             # LOC: 23370
grep -rE '#\[test\]|#\[tokio::test\]' \
  crates/harness/src | wc -l                         # tests: 312
grep -rE 'impl CweValidator for' \
  crates/harness/src/validation.rs | wc -l           # validators: 27

# build + tests
cargo test --locked

# a fresh, fully-instrumented run (protocol, App. F)
neurosploit run <target> \
  --capability-token <path> \
  --model <provider:model@version> \
  --votes <n> --seed <seed> \
  --config artifact/config.yaml \
  --audit artifact/audit.log
# integrity
shasum -a 256 <each artifact> > MANIFEST.sha256
shasum -a 256 MANIFEST.sha256   # bundle root
\end{lstlisting}

\section{The \texttt{supports} Decision Procedure}\label{app:supports}
\textbf{Input:} a claim $c$, its cited receipts $\mathrm{cite}(c)$, the engagement mode, and the target/action binding of each receipt.
\textbf{Output:} \textsf{Confirmed}, \textsf{Rejected}, or \textsf{NeedsReview}.
\textbf{Procedure.} (1) \emph{Binding}: each receipt's action hash must match a logged action against an in-scope target and postdate that action; a receipt that cannot be bound is dropped. (2) \emph{Class dispatch}: route to the per-CWE validator (Appendix~\ref{app:validators}); a class with no validator yields \textsf{NeedsReview} (never auto-\textsf{Confirmed}). (3) \emph{Composition}: all receipt \emph{roles} the class requires must be present (e.g., IDOR requires an owner-resource receipt \emph{and} a cross-identity receipt). (4) \emph{Contradiction}: if any receipt refutes the claim (e.g., the anonymous request was denied), return \textsf{Rejected}. (5) \emph{Verdict}: \textsf{Confirmed} iff roles present, bound, and the class predicate holds; else \textsf{NeedsReview}. The procedure is deterministic and model-free; it is the operational meaning of ``outside the model'' in P1.

\section{Deterministic Validator Catalog}\label{app:validators}
The 27 validators reference 31 CWE identifiers, including CWE-16, 22, 78, 79, 89, 113, 200, 209, 306, 307, 319, 347, 352, 384, 524, 525, 530, 548, 601, 611, 614, 639, 644, 650, 863, 915, 918, 942, 1004, 1021, 1336. Representative predicates: \textbf{SQLi (89)} a reproducible baseline-vs-attack differential; \textbf{XSS (79)} a browser executing a harness-chosen marker (proof by execution); \textbf{IDOR (639)} identity B reading identity A's resource with content-overlap confirmation, rejecting identical identities or generic login-page bodies; \textbf{SSRF (918)} an out-of-band callback carrying the marker or retrieval of a controlled internal resource; \textbf{missing authentication (306)} a protected resource served to an anonymous request with content match, rejecting the case where the anonymous request is denied.

\section{The Evidence Prosecutor}\label{app:prosecutor}
The prosecutor answers, in order: what was observed; which sentence exceeds it; what would make the impact factual; and whether the finding survives removing the impact sentence. It returns one of \{\textsf{supported\_finding}, \textsf{informational\_observation}, \textsf{insufficient\_evidence}, \textsf{invalid\_claim}\}; it cannot silently delete, and it cannot force-retain a claim with no security-relevant mechanic (which becomes \textsf{informational\_observation} or \textsf{insufficient\_evidence}). The finding is rebuilt from the returned minimal supported claim, decoupling the reduction (this component) from rejection (the vote and validators), which is what prevents a proven mechanic from being lost to an overstated headline.

\section{Severity Cap and Worked Example}\label{app:cvss}
Severity uses the CVSS v3.1 base metrics \cite{first2019cvss}; each metric value must cite a receipt. The demonstrated-impact ladder \eqref{eq:ladder} caps the base score:
\begin{center}
\small
\begin{tabular}{@{}lll@{}}
\toprule
Rung & Typical impact metrics & $\mathrm{cap}$ (base) \\
\midrule
reached & C:N/I:N/A:N, no data & $\le 6.9$ (Med) \\
read & C:L/H, I:N & $\le 8.9$ (High) \\
wrote & I:L/H & $\le 9.4$ \\
RCE & C:H/I:H/A:H & $\le 10.0$ (Crit) \\
crossed & scope S:C, C/I/A:H & $\le 10.0$ (Crit) \\
\bottomrule
\end{tabular}
\end{center}
\emph{Example.} A reflected SQL error on \texttt{login.aspx} with no data extraction: vector \texttt{AV:N/AC:L/PR:N/UI:N/S:U/C:N/I:N/A:N} yields a low base, and $\ell_d=$ reached caps at Med; the \emph{potential} score $\mathrm{sev}_p$ records C:H/I:H (a full extraction) so the un-demonstrated worst case is not lost. A \textsf{NeedsReview} metric leaves the finding at label-only severity with no computed base until resolved.

\section{Capability Token and Key Custody}\label{app:token}
Wire form: \texttt{ns-cap.v1.}\allowbreak\texttt{<b64url(payload)>.}\allowbreak\texttt{<b64url(mac)>}. Payload fields: \texttt{iss} (client key id), \texttt{aud} (harness instance), \texttt{eng} (engagement id), \texttt{op} (operator id), \texttt{scope} (hosts/CIDRs/URL prefixes), \texttt{ceiling} (max action kind, risk), \texttt{env}, \texttt{nbf}, \texttt{exp}, \texttt{nonce}. Verification follows Section~\ref{sec:framework}. \emph{Custody:} the current HMAC form uses a shared secret and proves holder-of-key only; to bind client and operator, an asymmetric signature (client-held Ed25519 private key, harness-held public key) is the recommended evolution, following the caveat model of Macaroons \cite{birgisson2014macaroons} and the registered-claims model of JWT \cite{rfc7519}. \emph{Replay:} \texttt{nonce} freshness plus short \texttt{exp}. \emph{Revocation:} a revocation list keyed by \texttt{eng}/\texttt{iss} and short lifetimes. \emph{Rotation:} \texttt{iss} key id selects the verifying key, enabling overlap during rotation.

\section{Audit Log Schema and Anchoring}\label{app:audit}
Each record carries \texttt{timestamp, agent, hypothesis, action, target, policy\_decision, tool, result, evidence\_hash, capability\_token}, and $h_i=H(r_i\|h_{i-1})$. \texttt{verify()} recomputes the chain and localizes the first break. Against a local adversary this detects removal, alteration, and reordering. Against a full-rewrite adversary, external anchoring is required: periodic signing of $h_i$ with a key outside the run, a trusted timestamp, a write-once store, or publication of roots to a transparency log \cite{rfc6962,crosby2009,schneier1999}. The acceptance test T-P5 exercises removal, alteration, reordering, evidence substitution, token swap, truncation, and whole-chain rebuild; the last passes only with anchoring enabled.

\section{Evaluation Protocol (to be executed)}\label{app:protocol}
This protocol specifies the measurements not yet performed; it is stated so results are comparable when produced. The deterministic P4 and P5 batteries below were partially executed already (Section~\ref{sec:accept}, Table~\ref{tab:accept}); the multi-target, ablation, and baseline measurements, which require live model runs, remain to be executed.
\textbf{Targets:} $\ge$ 3--5 applications with published ground truth---OWASP Juice Shop, WebGoat, DVWA (pinned versions), a deliberately-vulnerable API, and one authorized real scope---each with an enumerated expected-vulnerability list.
\textbf{Design:} $\ge 3$ runs per target with fixed model/version, temperature, vote count, seed, token budget, and concurrency, all recorded in the run config.
\textbf{Metrics:} true positives, false positives, false negatives against ground truth; precision, recall; cost, tokens, wall-clock; and, for P3, the CVSS delta between evidence-graded and by-class scoring.
\textbf{Ablation:} remove each deterministic validator, the vote, the prosecutor, and the P3 cap; run with and without P4; compare common logging vs.\ anchored P5---reporting the change in false-positive rate and cost attributable to each.
\textbf{Baselines:} NeuroSploit with P1--P5 disabled, a plain ReAct loop, a single-model configuration, and a conventional scanner on the same targets.
\textbf{Adversarial P4 (T-P4):} off-scope redirect, DNS rebinding, alternate IPv6/decimal/hex encodings, CNAME to external host, userinfo URL, over-broad wildcard, unauthorized port, mid-run DNS change, discovered subdomain, expired token, tampered token, in-session widening---each must be denied and logged.
\textbf{Adversarial P5 (T-P5):} record removed, altered, reordered; evidence substituted; token swapped; log truncated; whole chain rebuilt---each must be detected (the last with anchoring).

\section{Comparative Coding Protocol (Table~\ref{tab:related})}\label{app:table}
Each cell of Table~\ref{tab:related} was assigned from the analyzed artifact of each system, using fixed criteria. \textbf{Analyzed artifacts:} PentestGPT \cite{deng2023pentestgpt} (paper and public repository), Cochise \cite{happe2026cochise} (paper and 630-LOC release), MAPTA \cite{david2025mapta} (paper), PentestJudge \cite{caldwell2025pentestjudge} (paper), and NeuroSploit \cite{neurosploit2026} (source at the pinned commit, Table~\ref{tab:pin}). \textbf{Coding rule:} \cmark{}~= the artifact describes or implements a mechanism meeting the property's acceptance test (Section~\ref{sec:framework}); \pmark{}~= a partial or process-level mechanism that does not fully meet the test; \xmark{}~= the artifact makes no claim to the property. An \xmark{} records the absence of a claim in the analyzed artifact and is \emph{not} evidence of absence in code that was not reviewed. \textbf{Per-property predicate:} P1---an out-of-model check that a receipt supports a claim; P2---a documented rule that reduces rather than deletes an overstated finding; P3---a computed, evidence-graded severity (not by-class); P4---a signed, verifiable scope grant enforced as a ceiling; P5---an append-only, integrity-verifiable action log. \textbf{NeuroSploit assignments:} P1 \cmark{} (grounding and 27 validators, Section~\ref{sec:accept}); P2 \cmark{} (prosecutor with four typed outcomes); P3 \pmark{} (calculator and cap implemented, but CVSS vectors not populated in the archived run); P4 \pmark{} (token integrity enforced, primary-target check missing); P5 \pmark{} (hash chain present and tested, external anchoring design-only). A second independent coder is recommended before peer review; the criteria are published here so the coding is contestable.

\end{document}